\documentclass[sigconf,screen]{acmart}

\usepackage[utf8]{inputenc}
\usepackage{wrapfig}
\usepackage{graphicx}
\usepackage{amsmath}
\usepackage{amsthm}
\usepackage{booktabs}
\usepackage{algorithm}
\usepackage{multirow}
\usepackage{wrapfig}
\usepackage{marvosym}
\def\eps{{\epsilon}}

\def\mA{{\mathbf{A}}}

\def\mU{{\mathbf{V}}}

\def\vzero{{\bm{0}}}

\def\valpha{{\bm{\alpha}}}

\def\vphi{{\bm{\phi}}}

\def\vc{{\bm{c}}}
\def\vd{{\bm{d}}}

\def\vv{{\bm{v}}}

\def\vx{{\bm{x}}}

\def\vz{{\bm{z}}}

\def\mA{{\bm{A}}}

\def\mC{{\bm{C}}}
\def\mD{{\bm{D}}}

\def\mI{{\bm{I}}}

\def\mL{{\bm{L}}}

\def\mU{{\bm{U}}}
\def\mU{{\bm{U}}}

\DeclareMathAlphabet{\mathsfit}{\encodingdefault}{\sfdefault}{m}{sl}
\SetMathAlphabet{\mathsfit}{bold}{\encodingdefault}{\sfdefault}{bx}{n}

\def\gL{{\mathcal{L}}}

\def\gL{{\mathcal{L}}}

\usepackage{algorithm}
\usepackage{algpseudocode}

\def\gL{{\mathcal{L}}}

\newcommand{\firstone}[1]{\colorbox{red!15}{#1}}
\newcommand{\secondone}[1]{\colorbox{blue!15}{#1}}
\def\vzero{{\bm{0}}}

\def\vc{{\bm{c}}}
\def\vd{{\bm{d}}}

\def\vv{{\bm{v}}}

\def\vx{{\bm{x}}}

\def\vz{{\bm{z}}}

\def\valpha{{\boldsymbol{\alpha}}}

\def\vsigma{{\sigma}}

\usepackage{array}
\usepackage{array}
\usepackage{hyperref}
\usepackage{mathrsfs}
\usepackage{booktabs}
\usepackage{bm}
\usepackage{subcaption}
\usepackage{multirow}
\usepackage{tikz}
\usetikzlibrary{chains, positioning}

\usepackage{pifont}

\AtBeginDocument{%
  \providecommand\BibTeX{{%
    \normalfont B\kern-0.5em{\scshape i\kern-0.25em b}\kern-0.8em\TeX}}}

\copyrightyear{2025}
\acmYear{2025}
\setcopyright{acmlicensed}\acmConference[WSDM '25]{Proceedings of the Eighteenth ACM International Conference on Web Search and Data Mining}{March 10--14, 2025}{Hannover, Germany}
\acmBooktitle{Proceedings of the Eighteenth ACM International Conference on Web Search and Data Mining (WSDM '25), March 10--14, 2025, Hannover, Germany}
\acmDOI{10.1145/3701551.3703490}
\acmISBN{979-8-4007-1329-3/25/03}

\begin{document}

\title{S-Diff: An Anisotropic Diffusion Model for Collaborative Filtering in Spectral Domain}








\author{Rui Xia}
\affiliation{%
  \institution{
  Nanjing University of Aeronautics and Astronautics}
  \city{Nanjing}
  \country{China}}
\email{xiarui@nuaa.edu.cn}
\author{Yanhua Cheng}
\affiliation{%
  \institution{
  Kuaishou Technology}
  \city{Beijing}
  \country{China}}
\email{chengyanhua@kuaishou.com }
\author{Yongxiang Tang}
\affiliation{%
  \institution{
  Kuaishou Technology}
  \city{Beijing}
  \country{China}}
\email{tangyongxiang@kuaishou.com }
\author{Xiaocheng Liu}
\affiliation{%
  \institution{
  Kuaishou Technology}
  \city{Beijing}
  \country{China}}
\email{liuxiaocheng@kuaishou.com }
\author{Xialong Liu}
\affiliation{%
  \institution{
  Kuaishou Technology}
  \city{Beijing}
  \country{China}}
\email{liuxialong2007@sina.com }
\author{Lisong Wang}
\affiliation{%
  \institution{Nanjing University of Aeronautics and Astronautics}
  \city{Nanjing}
  \country{China}}
\email{wangls@nuaa.edu.cn}
\author{Peng Jiang}
\affiliation{%
  \institution{
  Kuaishou Technology}
  \city{Beijing}
  \country{China}}
\email{jp2006@139.com}

\renewcommand{\shortauthors}{Rui Xia et al.}


\begin{abstract}
Recovering user preferences from user-item interaction matrices is a key challenge in recommender systems. While diffusion models can sample and reconstruct preferences from latent distributions, they often fail to capture similar users' collective preferences effectively. Additionally, latent variables degrade into pure Gaussian noise during the forward process, lowering the signal-to-noise ratio, which in turn degrades performance. To address this, we propose S-Diff, inspired by graph-based collaborative filtering, better to utilize low-frequency components in the graph spectral domain. S-Diff maps user interaction vectors into the spectral domain and parameterizes diffusion noise to align with graph frequency. This anisotropic diffusion retains significant low-frequency components, preserving a high signal-to-noise ratio. S-Diff further employs a conditional denoising network to encode user interactions, recovering true preferences from noisy data. This method achieves strong results across multiple datasets.
\end{abstract}




\maketitle
\section{Introduction}
In recent years, generative methods, particularly diffusion models, have gained significant popularity in recommendation systems \cite{diffrec, xu2024difashion, li2024dimerec}. These methods capitalize on the potential of complex denoising networks and iterative sampling processes to achieve exceptional performance. By incorporating multimodal learning \cite{ma2024multimodal_diff, diffusion_for_audio}, graph representation learning \cite{jiang2024diffmm, ma2024multimodal_diff, graph-diff}, contrastive learning \cite{cContrastive24diffusion, difaug}, and negative sampling strategies \cite{negediff, liu2024preference}, these models consistently demonstrate valuable advantages across diverse scenarios, establishing them as a highly sought-after approach in the community.

Despite recent advancements, diffusion-based recommendation models still face several challenges. One critical limitation is that traditional diffusion models primarily depend on individual user interaction vectors as conditional inputs, failing to fully utilize the rich shared preference information across users in collaborative filtering, which diminishes the model's capability of interpretation and generalization \cite{cf-diff,graph-diff,cfdiff2}. Moreover, injecting large amounts of Gaussian noise into high-dimensional historical interaction vectors complicates the recovery process for the denoising decoder \cite{benedict2023recfusion,diffrec}, a challenge that cannot be overlooked. On the one hand, while some models have attempted to explicitly use collaborative information as conditional guidance for the denoising network (decoder) to improve the quality of recovery \cite{cf-diff}, the forward process (encoder) does not reflect collaborative signals, which to some extent leads to inconsistency between the encoding and decoding processes. On the other hand, some models \cite{diffrec, direct-diff} encode users' high-dimensional interactions into latent layers using clustering methods to reduce the difficulty of  recovering from noise distribution. However, it is still unclear whether these latent representations effectively capture the unique characteristics of collaborative filtering data.

\begin{figure}[!tb]
\centering
\caption{Graph-guided diffusion: By smoothing the interaction signals of users through neighboring item adjacencies, thereby turning them into noise, it is possible to train a denoising network to recover the original signals.
}
\includegraphics[width=0.38\textwidth]{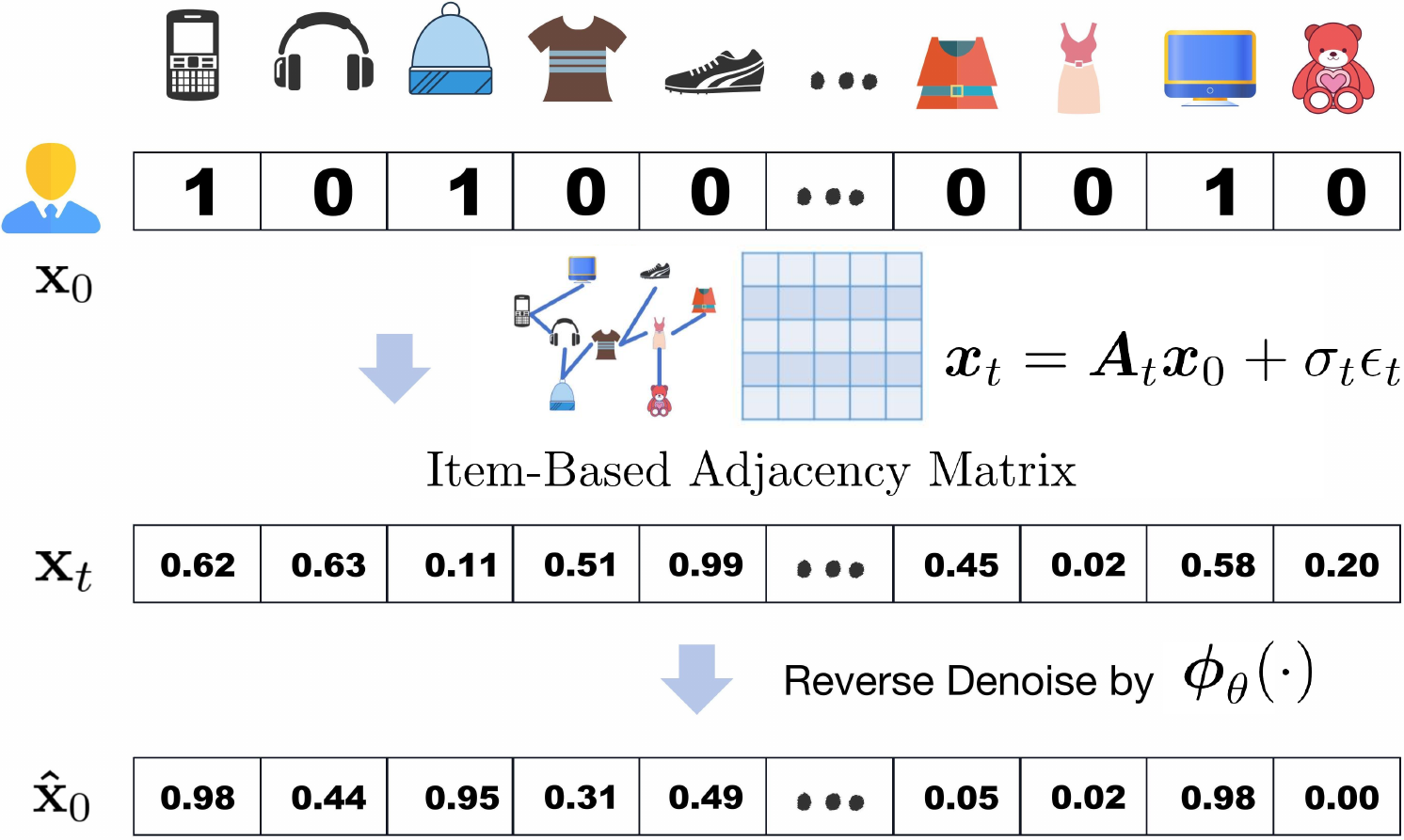}
\label{fig:intro}
\vspace{-5mm}
\end{figure}
Graph-based collaborative filtering methods \cite{BSPM, wang2019NGCF} offer a promising solution to the first challenge, as they are well-suited to capturing users' shared preferences by iteratively extracting interaction information from low to high orders, leading to notable performance improvements.
Notably, graph convolution is often seen as a destructive process, as it can gradually result in the "over-smoothing" of interaction signals. This reminds us of the probability diffusion guided by signal smoothing techniques applied in computer vision, such as image blurring and heat diffusion \cite{whang2022deblurring, cold}. These approaches have gone beyond the traditional Gaussian diffusion framework, leading to notable advancements in image and multimedia analysis.
Coincidentally, diffusion dynamics on graphs have become increasingly popular in recent years \cite{graphode,li2024generalizedgdiff}, which has also inspired our desire to combine this paradigm with probabilistic diffusion models, enabling user interaction signals to merge with global preferences during the forward destructive phase (see Fig. \ref{fig:intro}).

It is noteworthy that the success of graph collaborative filtering owes much to advances in graph signal processing, which emphasize the importance of leveraging graph smoothness to retain low-frequency information representing global preferences while filtering out high-frequency noise, thereby progressively smoothing interaction signals across nodes. Building on this, we introduce an anisotropic diffusion model in the item-based graph spectrum domain, where the noise scheduling parameter of the diffusion process is defined by the corresponding graph's eigenvalue coefficients.

Specifically, we correlate the noise scheduling parameters of the diffusion model with the eigenvalue coefficients of the graph spectrum. This approach preserves the low-frequency components during multi-step diffusion. Correspondingly, the reverse denoising process serves as a filter to remove the noise across different frequency components.

Additionally, to mitigate the challenging situation caused by excessive noise from forward diffusion, we propose flexible variance-preserving parameter settings that enhance the modulation of collaborative signals. This allows for a more accurate restoration of user preferences.
Specifically, our contribution lies in:
\begin{itemize} \item We introduce a forward diffusion process defined in the graph spectral domain, which effectively incorporates the global preferences of users.
\item We propose a parameterized approach to modulating the noise scale of different frequency components, and our analysis and experimental results on signal-to-noise ratio demonstrated the benefits of this setting.
\item Correspondingly, in the reverse process, we devise an element-wise fusion-based denoising module, and extensive experiments have validated the efficacy of our proposed method.
\end{itemize}

In Section \ref{Sec:back}, we introduce the background of diffusion models for collaborative filtering. In Section \ref{Sec:method}, we present the method proposed in this paper, transitioning from spatial diffusion models to those in the spectral domain. Section \ref{Sec:Experiment} records the results of the experiments, and finally, in Sections \ref{Sec:Related} and \ref{Sec:Conclusion}, we discuss the existing work and summarize the paper, respectively.

\section{Background}
\label{Sec:back}
\subsection{Preliminaries}
Collaborative filtering can be elegantly framed as an inverse problem, amenable to resolution through the application of generative models. We denote the set of users by $\mathcal{U}$ and the set of items by $\mathcal{I}$. For every user within $\mathcal{U}$, in the setting of collaborative filtering with implicit feedback, our dataset is encapsulated by a user-item interaction matrix $\boldsymbol{X} \in \{0,1\}^{| \mathcal{U} | \times | \mathcal{I} |}$. Each row $\boldsymbol{x}_u \in \{0,1\}^{| \mathcal{I} |}$ of this matrix represents the interaction vector for user $u$, where $\boldsymbol{x}_{u,i} = 1$ signifies an interaction between user $u$ and item $i$, and $\boldsymbol{x}_{u,i} = 0$ indicates the absence of such an interaction. For the sake of simplicity, we will henceforth refer to the user's historical interaction vector as $\boldsymbol{x}$. Our primary goal is to predict a score vector $\hat{\boldsymbol{x}} \in \mathbb{R}^{| \mathcal{I} |}$, which serves to generate potential preference scores (or probabilities) for all items in $\mathcal{I}$ for a specified user. Subsequently, we aim to recommend the top $K$ items that are most likely to align with and fulfill the user's preferences and needs.

\vspace{+2mm}
\noindent \textbf{Item-based Graph.}
The interaction matrix $\boldsymbol{X}$ serves as a bridge, delineating the adjacency relationships between user nodes and item nodes within the context of a user-item bipartite graph. To facilitate the smoothing of interaction vectors, it becomes imperative to quantify the similarity between any two item nodes. To this end, we introduce $\boldsymbol{D}_{\boldsymbol{U}} = \operatorname{diagMat}(\boldsymbol{X} \boldsymbol{1})$ and $\boldsymbol{D}_{\mathcal{I}} = \operatorname{diagMat}(\boldsymbol{X}^{\top} \boldsymbol{1})$, which respectively represent the degree matrices for users and items. Besides, we define the normalized similarity adjacency matrix for item-to-item relationships as $\boldsymbol{A} = \tilde{\boldsymbol{X}}^{\top} \tilde{\boldsymbol{X}}$, where $\tilde{\boldsymbol{X}} = \boldsymbol{D}_{\mathcal{U}}^{-\frac{1}{2}} \boldsymbol{X} \boldsymbol{D}_{\mathcal{I}}^{-\frac{1}{2}}$. This definition enables us to construct the Laplacian operator $\boldsymbol{L} = \boldsymbol{I} - \boldsymbol{A}$, which plays a pivotal role in graph spectral theory and is instrumental in our subsequent analysis and algorithm design.

\vspace{+2mm}
\noindent \textbf{Graph Fourier Transform (GFT).} Moreover, given that the Laplacian matrix $\boldsymbol{L}$ is semi-positive definite and normalized, it can be further decomposed via its eigenvalue decomposition: $\boldsymbol{L} = \boldsymbol{U}^T \boldsymbol{\Lambda} \boldsymbol{U}$. Here, $\boldsymbol{\Lambda} = \operatorname{diag}\{\lambda_1, \ldots, \lambda_{|\mathcal{I}|}\}$ is a diagonal matrix composed of the eigenvalues of $\boldsymbol{L}$,  and $\boldsymbol{U} = [\boldsymbol{u}_1, \ldots, \boldsymbol{u}_{|\mathcal{V}|}]$ represents a set of corresponding normalized orthogonal eigenvectors. We have the graph Fourier transform (GFT) \cite{gft} $\boldsymbol{v} = \boldsymbol{U} \boldsymbol{x}$, which maps the graph signal $\boldsymbol{x}$ into the graph spectral domain. In this case, each component $v^{(i)}$ of the vector $\vv$ represents the mapping of the corresponding spatial signal $\vx$ onto the eigenvalue $\lambda_i$.

\subsection{Ordinary DDPMs}

Representing user interaction vectors as a data distribution that is progressively noised over time steps to become increasingly obfuscated, and then reversing the process to sample and recover from it, forms the basis of generative models \cite{ddpm,diffrec}.

For simplicity, we denote the user's interaction vector $\vx$ as $\vx_0$, representing the interaction at the initial time step. Given an initial sample $\vx_0$ drawn from the initial user data distribution $q(\vx_0)$, the forward process of a classical Gaussian diffusion is defined by a sequence of increasingly noisy random variables that deviate from $\vx_0$ as, 
\begin{equation}
\label{eq:diff}
\boldsymbol{x}_t = \alpha_t \vx_{0} + \sigma_t \bm{\eps}_t, \quad \bm{\eps}_t \sim \mathcal{N}(\bm{0}, \bm{I}).
\end{equation}
By setting $\alpha_T$ close to $0$, $q\left(\boldsymbol{x}_T\right)$ converges to $\mathcal{N}(\mathbf{0}, \boldsymbol{I})$, we are able to sample data $\boldsymbol{x}_0$ by using a standard Gaussian prior and a learned inference model $p_\theta\left(\boldsymbol{x}_{t-1} \mid \boldsymbol{x}_t\right)$. The inference model $p_\theta$ is parameterized as a Gaussian distribution with predicted mean and time-dependent variance scale $\sigma_t^2$, and data can be sampled through sequential denoising, i.e., $p_\theta\left(\boldsymbol{x}_0\right)= \prod_{t=1}^T p_\theta\left(\boldsymbol{x}_{t-1} \mid \boldsymbol{x}_t\right)$.

Specifically, we focus on conditional guided diffusion models \cite{conditionaldiff}, that is, to recover the true preferences of users through their historical interactions $\vc$ as a condition. To estimate the conditional distribution $q(\vx_0|\vc)$, the reverse process can be defined as a parameterized hierarchical model: \begin{equation}
\small
p_{\theta}\left(\vx_{1: T} \mid \vx_0, \vc\right)=p_{\theta}\left(\vx_T \mid \vx_0, \vc\right) \prod_{t=1}^{T} p_{\theta}\left(\vx_{t-1} \mid\vx_{t}, \vx_0, \vc\right),
\end{equation} where we set
$p(\vx_T|\vx_0,\vc) = \mathcal{N}(\vx_T;\bm{0}, \bm{I})$ to guide an ELBO objective function based on $\vc$ for a conditional diffusion model.

We also set $p_{\theta}(\vx_{t-1}|\vx_t,\vc) = q(\vx_{t-1}|\vx_t, \vx_0 = \vphi_\theta(\vx_t, \vc, t))$ for $t=T,\dots,0$, as well as $p_{\theta}(\vx_0|\vx_1,\vc) = \mathcal{N}(\vx_0;\vphi_\theta(\vx_1, \vc, 1), \delta^2 \bm{I})$, where $\delta$ is the variance constant. The remaining task is to learn a neural denoiser ${\vphi}_{\theta}$ to predict the mean by maximum likelihood estimation.

During the sampling process, we start from random noise $\vx_T \sim p(\vx_T)$ and iteratively refine the noise latent variable $\vx_t$ by sampling $\vx_{t-1} \sim p_{\theta}(\vx_{t-1}|\vx_t, \vc)$ for all $t = T, \dots, 0$, ultimately obtaining a final denoised user preference vector $\hat{\vx}_{0}$.

\vspace{+2mm}
\noindent \textbf {Notation.} To avoid confusion, it is noteworthy that in this paper, the normalized time parameter $t = \tau * t_k / T \in [0,\tau]$ represents a conceptual time step. Here, $t_k$ denotes the actual discrete time steps commonly utilized in diffusion models and $\tau$ is a hyper parameter deciding the actual latency of the diffusion process. Besides, the bold font indicates a vector (or matrix), while the non-bold font represents a scalar, such as $\boldsymbol{\alpha}$ and its $i$-th element, $\alpha_i$.

\section{Spectral Diffusion Framework in Recommendation Systems}
\label{Sec:method}
\noindent \textbf{Organization.} In Sec. \ref{3.2}, we introduce a  approach utilizing graph differential equations to guide a forward diffusion process. In Sec. \ref{3.2puyu}, we generalize a unified spectral domain diffusion paradigm. Furthermore, we discuss the property and time complexity of the spectral domain diffusion paradigm in Sec. \ref{noise_scale}. In Sec. \ref{3.5}, we will outline the procedure for reverse sampling to facilitate recovery, as well as the denoising network employed in the process.

\label{3.1}
\begin{figure*}[!tb]
\centering
\caption{Spectral Domain Diffusion Models with Anisotropic Noise:
We perform an eigenvalue decomposition on the item-based Laplacian matrix, yielding an orthogonal matrix $\mU^{\top}$ that maps user interaction vectors $\vx$ into the spectral domain as $\vv$, i.e., the Graph Fourier Transform (GFT). We then introduce anisotropic noise into this spectral representation by leveraging the eigenvalues' varying influences. Afterward, we recover the users' original interaction signals from the noisy vector $\vv_{T}$ and use the Graph Inverse Fourier Transform (GIFT) to map the decoded signals $\hat{\vv}_0$ back to the spatial domain as $\hat{\vx}_u$.}
\includegraphics[width=0.9\textwidth]{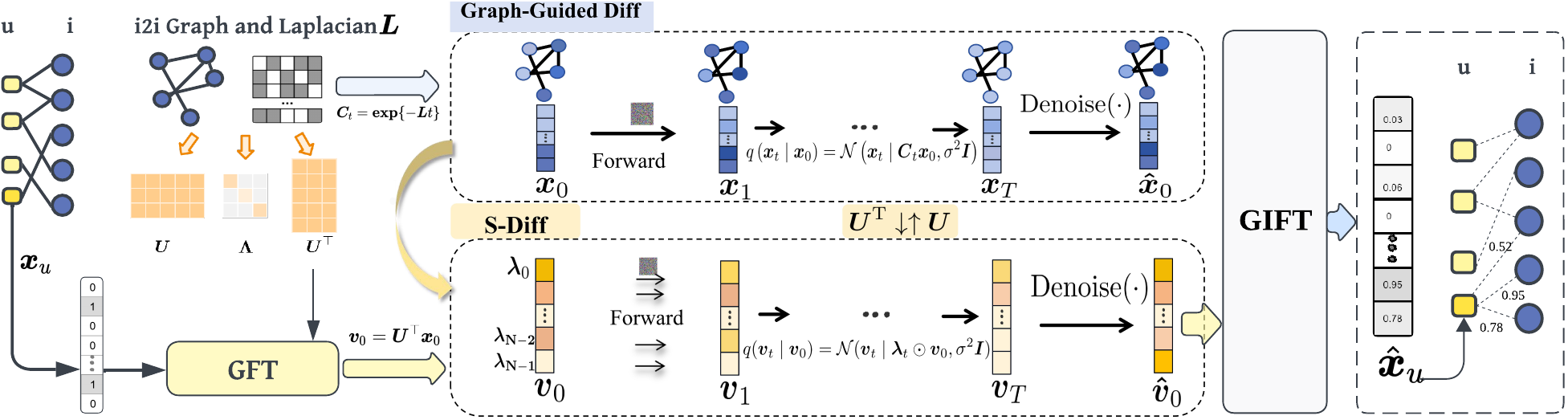}
\vspace{-3mm}
\end{figure*}

\subsection{Graph-Guided Forward Diffusion}
\label{3.2}

We revisit the form of equation (\ref{eq:diff}), inspired by the successes of cold diffusion and diffusion models based on ordinary differential equations (ODEs) across various domains \cite{cold, heatinverse2,whang2022deblurring,hagemann2023generalized,Hoogeboom2022EquivariantDF, graphode,graphpde}. Indeed, the forward process we construct can be generalized as matrix multiplication followed by additive noise. For instance, in the resolution of image degradation problems, matrix multiplication is instantiated as image blurring or masking to adhere to the prior of real-world problems \cite{ren2023multiscale}. When we have precise knowledge about certain properties of the problem to be solved, such a setup can aid in recovery. Inspired by this paradigm, our work slightly deviates from the paradigm of adding noise in scalar form, as seen in equation (\ref{eq:diff}), and instead considers the study of a degradation process in the following form:
\begin{gather}
    \vx_t = \mC_t \vx_0+ \sigma_t \bm{\eps}_t, 
    \label{eq:general_diffusion_form}
\end{gather}
$\text {where} \quad \boldsymbol{C}_t: \mathbb{R}^{|\mathcal{I}|} \rightarrow \mathbb{R}^{|\mathcal{I}|}  \quad 
\text {is a deterministic linear operator}$ that controls the degeneration in forward-diffusion, and $\sigma_t$ is the variance constant that controls the level of noise. 

As mentioned, we aspire for this degradation operator to align closely with the practices of collaborative filtering tasks. The diffusion of the graph heat kernel \cite{li2024generalizedgdiff,icadiff,lin2024graph} guided by the Laplacian matrix $\mL$ appears promising, as we have chosen the item-item adjacency matrix $\mA$. The heat conduction differential equation that unfolds along this matrix implies the smoothing of user interaction signals, which in turn reflects the homophilous preferences of users when clicking items. Specifically,  graph heat diffusion introduce  a differential equation  on the graph, where the greater the difference between two nodes, the faster the derivative smooths it out \cite{li2024generalizedgdiff},
\begin{gather}
\small
    \frac{d \vx^{(t)}_{i}}{dt} = \sum_{j \in \mathcal{N}(i)} { \mA_{ij}} \left( \boldsymbol{x}_j^{(t)} - \boldsymbol{x}_i^{(t)} \right)= \sum_{j \in \mathcal{N}(i)} { \mA_{ij}}  \boldsymbol{x}_j^{(t)} - \boldsymbol{x}_i^{(t)} .
    \label{graph_diff}
\end{gather}
This equation models the change of the signal $x_i^{(t)}$ at node $i$ over time $t$. Since the item-item graph's Laplacian matrix is $\mL=\mI-\mA$, the differential equation of the graph heat diffusion \cite{icadiff}  can be stated as:
\begin{equation}
\frac{d \boldsymbol{x}_t}{d t}=-  \boldsymbol{L} \boldsymbol{x}_t,
\label{graph_heat}
\end{equation}
where $\vx_t$ is the vector of node values at time $t$, then we proceed to characterize the time decay operator associated with the Laplacian matrix $\boldsymbol{L}$. Given the initial value $\vx_0$, we have the following when integrating on both sides in Eq. (\ref{graph_heat})
\begin{equation}
\boldsymbol{x}_t  = e^{ -\boldsymbol{L} t} \boldsymbol{x}_0.
\label{eq:Ct_heat}
\end{equation}

The diffusion equation can be stated using the $\mC_t(\cdot)$ satisfied
\begin{equation}
    \mC_t \vx_0= e^{ -\boldsymbol{L} t} \vx_0.
    \label{7}
\end{equation}

Eqs. (\ref{eq:Ct_heat}) and (\ref{7}) show a non-stochastic diffusion process defined by item-item graph. Since the powerful effect of the stochastic diffusion methods such as DDPMs shown in \cite{ddpm,ddpm++}, we introduce a Gaussian process $\vz_t=\sigma_t \bm{\eps}_t$ to serve as noise of our hybrid diffusion process on the graph. Here, scalar sequence $\sigma_t$ increases monotonically and  $\vz_t$ satisfies $\vz_{t_2}-\vz_{t_1} \sim \mathcal{N}(0, (\sigma_{t_2}^2 - \sigma_{t_1}^2)\boldsymbol{I})$ for any $t_2>t_1>0$, so that $\bm{\eps}_t \sim \mathcal{N}(0,\boldsymbol{I})$. Now we have the forward process guided by both the diffusion process on the item-item graph and the stochastic Gaussian process:$ \boldsymbol{x}_t = \boldsymbol{C}_t \boldsymbol{x}_0 + \sigma_t \bm{\eps}_t$, which also leads to the distribution of the forward process given by:
\begin{equation}
q\left(\boldsymbol{x}_t \mid \boldsymbol{x}_0\right)=\mathcal{N}\left(\boldsymbol{x}_t \mid \mu = \boldsymbol{C}_t \boldsymbol{x}_0, \Sigma = \sigma^2_t \boldsymbol{I}\right) .
\end{equation}
\begin{figure}[!tb]
\centering
\caption{We compare the SNR of the spectral diffusion model across datasets with traditional DDPM using Gaussian noise. Adding eigenvalue-related noise in the spectral domain prevents SNR degradation into white noise.}
\includegraphics[width=0.45\textwidth]{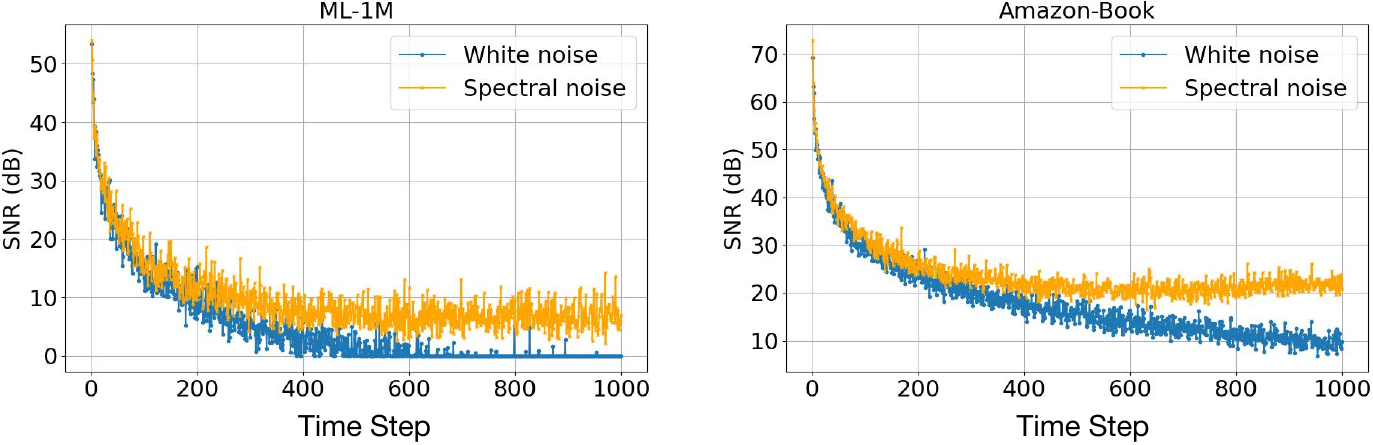}
\label{fig:snr}
\vspace{-4mm}
\end{figure}
\vspace{-2mm}
\subsection{Spectral Diffusion}

\label{3.2puyu}

In this section, we set $({ -\boldsymbol{L} t})=\mU \boldsymbol{\mD_t}\mU^{\mathrm{T}}$, where $\mU^{\mathrm{T}}$ denotes the orthonormal basis of eigenvectors obtained from matrix decomposition\footnote{In practical implementation, we use the Lanczos method to compute the eigenvectors corresponding to the top 200 eigenvalues for the feature space mapping}, in which $\mD_t$ is defined by a diagonal matrix whose entries are the eigenvalues arranged as follows:$\mD_t = \operatorname{diag}\{-t d_1, \ldots, -t d_{|\mathcal{I}|}\}$

where ${-t} \cdot d$ is the corresponding eigenvalue of $-\mL t$.  According to the definition of the exponential equation for diagonalizable matrices \cite{leonard1996matrix}, we have $\mC_t = e^{ -\boldsymbol{L} t} = \mU \boldsymbol{\Lambda}_t\mU^{\mathrm{T}} $ satisfied:
$$
\boldsymbol{\Lambda}_t = \operatorname{diag}\left(e^{-t \cdot d_1}, e^{-t \cdot d_2}, \ldots, e^{-t \cdot d_{|\mathcal{I}|}} \right).
$$

Under the spectral transform $\vv_t = \mU^{\mathrm{T}} \vx_t$, we can express the diffusion process for $\vv_t$ in the spectral domain as:
\begin{equation}
\boldsymbol{v}_t=\boldsymbol{\lambda}_t \odot \boldsymbol{v}_0+\boldsymbol{\sigma}_t \boldsymbol{v}_{\epsilon, t} \quad \text { where } \quad \boldsymbol{v}_0=\boldsymbol{U}^{\mathrm{T}} \boldsymbol{x}_0 \text { and } \boldsymbol{v}_{\epsilon, t}=\boldsymbol{U}^{\mathrm{T}} \boldsymbol{\epsilon}_t.
    \label{eq:jump_freq}
\end{equation}
Here, $\vv_{0} = \mU^{\mathrm{T}} \vx_0$ is the frequency response of $\vx_0$, $\boldsymbol{\lambda}_t = [e^{{-t} \cdot d_1}, e^{{-t} \cdot d_2}, \\ e^{{-t} \cdot d_3} \cdots,e^{{-t} \cdot d_{|\mathcal{I}|}}]\in \mathbb{R}^{{|\mathcal{I}|}}$ is the vector formed by the diagonal eigenvalues of $\boldsymbol{D}_t$, and the Hadamard product is performed element-wise.

According to the prior properties of the Gaussian distribution, the frequency domain mapping of Gaussian noise $\boldsymbol{v}_{\epsilon, t}$ in Equation ($\ref{eq:jump_freq}$) can be re-substituted with a Gaussian noise vector $\boldsymbol{\epsilon}_t$, and we let ${\sigma}_t$ be its constant component such that we have noise scale vector $\boldsymbol{\sigma}_{t} = \boldsymbol{1} \cdot {\sigma}_t$. 

Equation (\ref{eq:jump_freq}) indicates that the marginal distribution of frequency $\vv_t$ decomposes entirely over its scalar elements $v^{(i)}_t$. Similarly, the reverse spectral diffusion model $p_{\theta}(\vv_{t-1} | \vv_t)$ also decomposes entirely. Consequently, we can equivalently describe the scalar form of the diffusion process for each eigenvalue dimension $i$ as follows:
\begin{equation}
\small
q(v^{(i)}_t | v^{(i)}_0) =  \mathcal{N}(v^{(i)}_t | \lambda^{(i)}_t v^{(i)}_0, ({\sigma_t^i})^2) =  \mathcal{N}(v^{(i)}_t | e^{-t{d}_{i} }v^{(i)}_0, ({\sigma_t^i})^2) , \label{eq:heat_scalar}
\end{equation}

where $v^{(i)}_t = {\lambda}^{(i)}_t v^{(i)}_0 + \sigma_t \epsilon_t, \text{ with }  \epsilon_t\sim\mathcal{N}(0,I).$ This means that performing spatial diffusion guided by the Laplacian matrix can equivalently be defined as a relatively standard Gaussian diffusion process. This diffusion process, defined in the frequency space $\mU$, features noise that fluctuates along the amplitude of each eigenvalue, making it anisotropic.
Through the equivalence to Gaussian diffusion, we similarly define the spectral diffusion process and the noise scheduling coefficients as follows:
\begin{equation}
    q(\vv_t | \vv_{0}) = \mathcal{N}(\vv | \valpha_{t} \odot \vv_{0}, \boldsymbol{\sigma}^2_{t}\mI).
    \label{eq:rissanen_markov}
\end{equation}
Substituting the choices from \cite{cold}, we have:
\label{vpdiff}
\begin{equation}
\valpha_{t} = \boldsymbol{\lambda}_t \hspace{0.5cm} \text{and} \hspace{0.5cm} \boldsymbol{\sigma}_{t}^2 = \boldsymbol{1} -  \boldsymbol{\lambda}_t ^2,
\label{spectralalpha}
\end{equation}
to construct a variance-preserving diffusion model.

If $\lambda^{(i)}_t = e^{-td_i}$ is chosen such that it maintains lower values at higher frequencies $d_i$, this is because the negative exponent of $e$ is decreasing. Then $\vsigma^{(i)}_{t}$ will add \textit{more noise} at each time step to the \textit{higher frequency eigenvectors}. Consequently, compared to the diffusion model with standard Gaussian noise, the anisotropic noise introduces noise with specific frequencies (particularly the abnormal high-frequency components) and relatively retains the low-frequency components, which is very similar to the over-smoothing \cite{howpowerful,wei2019mmgcn} state in graph representation learning.

\vspace{+2mm}
\noindent \textbf{Boundedness Property.} Continuing with the previous notation, where $\boldsymbol{\alpha}_t$ and $\boldsymbol{\sigma}_t$ represent the noise schedule. For a random variable, the signal-to-noise ratio (SNR) is defined as the ratio of the mean squared value to the variance. Given that the mean of $\boldsymbol{v}_0$ and $\boldsymbol{v}_t$ is obviously $\boldsymbol{\alpha}_t \mathbb{E}\left[\boldsymbol{v}_0\right]$, and the variance is $\boldsymbol{\sigma}_t^2$, the SNR is thus $\frac{\boldsymbol{\alpha}_t^2}{\boldsymbol{\sigma}_t^2} \mathbb{E}\left[\boldsymbol{v}_0\right]^2$. Since we are always discussing under the condition of given $\boldsymbol{v}_0$, we can also simply say that the SNR is $SNR(t) = \frac{\boldsymbol{\alpha}_t^2}{\boldsymbol{\sigma}_t^2}$, in the commonly used DDPMs, it satisfies $\boldsymbol{\alpha}_0=\boldsymbol{\sigma}_T=\boldsymbol{1}, \boldsymbol{\alpha}_T=\boldsymbol{\sigma}_0=\boldsymbol{0}$, and in addition, they generally have additional constraints, such as in DDPMs, it is usually $\boldsymbol{\alpha}_t^2+\boldsymbol{\sigma}_t^2=1$. This constraint is also adopted in this paper.

In this case, as $t$ becomes sufficiently large, the SNR will approach zero. This paper state the case of anisotropic diffusion in the spectral domain. Since all eigenvalues $\vd = [d_1,d_2,\ldots,d_{|\mathcal{I}|}]$ of the normalized Laplacian matrix $\boldsymbol{L}$ are between [0, 2], and $t \in [0,\tau]$, the eigenvalues of $\boldsymbol{C}_t$ will be within the range [$e^{-2\tau}$, $1$]. This means that we consider: $SNR(t) = \frac{\boldsymbol{\alpha}_t^2}{\boldsymbol{\sigma}_t^2} = \frac{\boldsymbol{\alpha}_t^2}{1-\boldsymbol{\alpha}_t^2}$ as the range of this function is greater than $\frac{\left(e^{-2\tau}\right)^2}{1-\left(e^{-2\tau}\right)^2}$, which is to say that the SNR has a good lower bound in S-Diff.

\begin{algorithm}[!t]
\begin{algorithmic}
\small
\caption{ Training} 
\label{alg:training}
    \Require $ \vphi_{\theta},  \boldsymbol{ \sigma_t},\boldsymbol{\alpha_t}, \mU,  T$
    \Repeat
    \State $\text{Sample } u \text { from } \text{User Set} \text { and let } \boldsymbol{x}_0 \leftarrow \boldsymbol{x}_u, \boldsymbol{c} \leftarrow Mask(\boldsymbol{x}_u) \text {. }$
    \State $\vv_0 = \mU^{T}{\vx_0} $
    \State Sample $t \sim \mathcal{U}(0, 1)$ and $\boldsymbol{v}_t \sim \mathcal{N}(\vv_t | \valpha_{t} \odot \vv_{0}, \boldsymbol{ \sigma}^2_{t}\mI) $
    \State $\mathbf{c} \leftarrow \varnothing$ with probability $p_{\text {uncond }}=0.02$
      \State $\hat \vv_0 = \vphi_{\theta}(\vv_t, \boldsymbol{U}^{T}\vc,t) $
      \State   Take a gradient step on $\nabla_
                \theta\|\hat \vv_0 - \vv_0 \|^2$.
    \Until{convergence}
    \State \textbf{Output:} Denoiser parameters $\theta$.
\end{algorithmic}
\end{algorithm}

\begin{figure}[!t]
\vspace{-4mm}
\centering
\caption{Instantiation of Denoiser $\vphi_{\theta}(\cdot)$}
\includegraphics[width=0.25\textwidth]{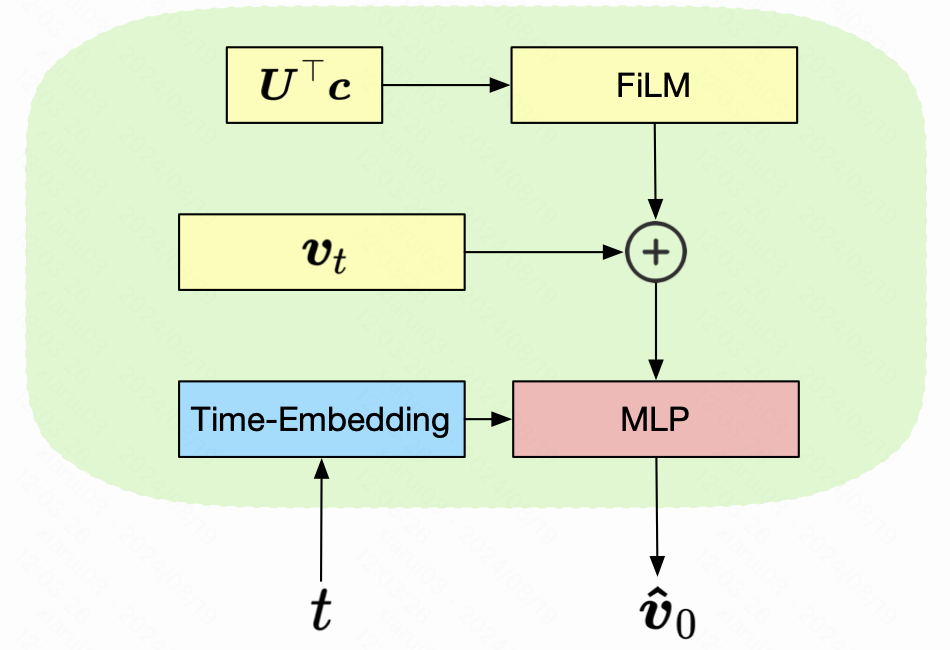}
\label{deoiser}
\vspace{-7mm}
\end{figure}

\vspace{+2mm}
\noindent \textbf{Case Study.} 
 As illustrated in Figure \ref{fig:snr}, under the constraints of bounded anisotropic noise, our spectral domain diffusion process, in contrast to traditional diffusion models, even after 1000 time steps, still retains a discernible signal-to-noise ratio for interaction signals, unlike the complete degeneration into meaningless Gaussian noise observed in traditional models.

\vspace{+2mm}
\noindent \textbf{Parameter Control.} In equation (\ref{spectralalpha}), we have established that $\boldsymbol{\alpha}_t$ is related to $\left(\boldsymbol{\lambda}_t\right)$, forming an anisotropic diffusion paradigm that relies on the eigenvalues of the graph. However, this means that the some eigenvalues may not benefit significantly from the diffusion paradigm. In fact, their recovery may be hindered by high levels of noise. In the case of recommendation data, it is unnecessary to add data to pure noise \cite{benedict2023recfusion, ber}. Instead, a smaller number of time steps, typically ranging from 3 to 5 steps \cite{diffrec}, is used to create a weakly noisy state for recovery. Fortunately, nerual diffusion process benefits from the precise parameterization of time step \cite{daras2022soft,ddrm}. Following the practice in \cite{hoogeboom2022blurring}, we introduce two hyperparameters, namely $\alpha_{min}$ and $\sigma_{max}$.

The hyperparameter $\alpha_{min}$ controls the retention of the minimum frequency. This prevents us from ignoring a sample when its feature value is too small. 
Additionally, the hyperparameter $\sigma_{max}$ controls the maximum noise ratio. It ensures that within the limited number of time steps, the state vector $\vx_t$ does not become dominated by pure noise.
In summary, by setting appropriate values for $\alpha_{min}$ and $\sigma_{max}$, we can fine-tune the noise ratio in the diffusion process, which ensures that the diffusion paradigm effectively captures the signal in the data while mitigating the negative impact of noise.
\begin{align} 
\small
\boldsymbol{\alpha}_t &= \left(1-\alpha_{\min }\right) \cdot \left(\boldsymbol{\lambda}_t \right)+\alpha_{\min}, \\ \boldsymbol{\sigma_t} &= \text{Min} \left(\sqrt{1-\boldsymbol{\lambda_t}^2}, \sigma_{\max }\right).
\label{diff_para}
\end{align}

\noindent \textbf{Time Complexity.} Compared to the regular diffusion model, add spectral diffusion operates under the decomposition dimension $\mathbb{R}^K$, where $K$ is the approximately truncated dimension, such as the first 200 eigenvalues, utilizing the Lanczos \cite{cullum2002lanczos} method for matrix decomposition with time complexity of $\mathcal{O}(K|\mathcal{I}|m)$, where $|\mathcal{I}|$represents the number of nodes in the item-similarity matrix, $m=10$ represents the number of iterations in the approximate matrix decomposition process.
 
 \label{noise_scale}

\begin{algorithm}[!t]
\begin{algorithmic}
\small
\caption{Sampling} 
\label{alg:naive_sampler}
    \Require $ \vphi_{\theta},  \boldsymbol{\sigma_t},\boldsymbol{\alpha_t}, \mU,  T, s$
    \State $\text{Sample } u \text { from } \text{User Set} \text { and let } \boldsymbol{c} \leftarrow \boldsymbol{x}_u \text {. }$
    \State $\vv_0 = \mU^{T}{\vc} $
    \State  Obtain $\vv_T $ via $\boldsymbol{v}_T =\boldsymbol{\alpha}_T \odot \boldsymbol{v}_0+\boldsymbol{ \sigma}_t \boldsymbol{\epsilon}_{T} $
    \For{$t \text{ in} \left\{\frac{T}{T}, \ldots, \frac{1}{T}\right\} \,\,$}
      \State $\hat \vv_0 = (1-s)\vphi_{\theta}(\vv_t, \boldsymbol{U}^{T}\vc,t) + s \vphi_{\theta}(\vv_t, \boldsymbol{0},t)$
      \State $\epsilon_t \sim \mathcal N(\vzero, \mI)$ 
      \State $\vv_{t -1} \gets \boldsymbol{\alpha}_{t - 1} \hat \vv_0 + \boldsymbol{\sigma}_{t - 1}\epsilon_t$ 
    \EndFor
    \State \textbf{return} $\hat{\vx}_0 = \mU \hat \vv_0$
\end{algorithmic}
\end{algorithm}

\vspace{-3mm}
\subsection{ Conditional Reverse and Optimization}
\label{3.5}
\noindent \textbf{Loss function.}
\label{loss}
The loss for a conditional diffusion model is the optimization of a lower bound on the negative log-likelihood of the generated data (ELBO). Specifically, we take the user's initial interaction history as the conditioning vector $\vc$, and step by step, minimizing the KL divergence between the predicted reverse distribution $p_\theta(\boldsymbol{x}_0|\vc)$ and the forward noise distribution $q(\boldsymbol{x}_{1: T}|\vc)$ under this condition. We have the decomposed form of the ELBO as:
\begin{equation}
\scriptsize
\begin{aligned}
\mathcal{L}_{\text{ELBO}} &= \mathbb{E}\left[-\log p_{\boldsymbol{\theta}}(\boldsymbol{x}_0 \mid \vc)\right] \\
&\leq \mathbb{E}\left[-\log \frac{p_{\boldsymbol{\theta}}(\mathbf{x}_{0: T})}{q(\boldsymbol{x}_{1: T} \mid \boldsymbol{x}_0)}\right]\\
& =\mathbb{E}\left[\sum_{t>1} D_{K L}\left(q\left(\boldsymbol{x}_{t-1} \mid \boldsymbol{x}_t, \boldsymbol{x}_0, \vc\right) \| p_{\boldsymbol{\theta}}\left(\boldsymbol{x}_{t-1} \mid \boldsymbol{x}_t, \vc\right)\right)\right] \\
&\quad -\mathbb{E}_{q\left(\boldsymbol{x}_1 \mid \boldsymbol{x}_0\right)} \left[\log p_{\boldsymbol{\theta}}(\boldsymbol{x}_0 \mid \boldsymbol{x}_1, \vc) \right].
\end{aligned}
\end{equation}

We primarily focus on the general step for sampling and recovery at the intermediate $t$ steps, denoted as $\gL_t$. Specifically, we utilize a neural network ${\vphi}_\theta(\cdot)$, which takes the conditional vector and the noise distribution as inputs, aiming to minimize the discrepancy between the network's output and the user's true interaction vector as the optimization objective.
 As for the recovery model in the spectral domain, its optimization objective satisfies:
\begin{gather}
    \mathcal{L}_t = \mathbb{E}_{(\vv_0, \vv_t) \sim q_0(\vv_0) q_t(\vv_t|\vv_0)}\left[ \left|\left| \left(\vphi_{\theta}(\vv_t, \boldsymbol{U}^{T}\vc,t) - \vv_0\right)\right|\right|^2 \right],  
\end{gather}
$\text{ where } \vv_0 = \mU^T  \vx_0$ and  $\vc$ represents the conditional vector (For example, in a real-world scenario, it is a sequence of past interactions of the user). During the sampling process, we use $\vphi_{\theta}$ for T-step predictions.


\vspace{+2mm}
\noindent \textbf{Classifier-Free Guidance.}
To emphasize the effectiveness of the conditional information (user historical interaction information) $\vc$, we adopt a classifier-free guidance method \cite{ho2022classifier}. This method simultaneously trains conditional and unconditional diffusion models without the need for classifier guidance on noise signals. Unlike classification-guided techniques, a classifier-free guidance method allows for a balance between sample quality and diversity during inference by adjusting the weighting between conditional and unconditional sampling. Specifically, during training, we maintain a 2\% batch dropout for conditional training, i.e., $(\left(\vphi_{\theta}(\vv_t, \mathbf{0} ,t)\right)$.

\vspace{+2mm}
\noindent \textbf{Instantiation of Denoiser.}
To ensure consistency with the element-wise anisotropic diffusion in the forward process, we learn the element-wise fusion of the conditional vector and the noise vector during the reverse conditioning recovery using Feature-wise Linear Modulation (FiLM) layer \cite{film}. Specifically, we utilize the FiLM (Feature-wise Linear Modulation) learning functions $f$ and $h$, which output $\gamma_i$ and $\beta_i$ as functions of the element-wise weight of the input $(\boldsymbol{U}^{\top} \boldsymbol{c})_i$ and the bias term:
\begin{equation}
\gamma_i=f\left((\boldsymbol{U}^{\top} \boldsymbol{c})_i\right), \quad \beta_i=h\left((\boldsymbol{U}^{\top} \boldsymbol{c})_i\right),
\vspace{-2mm}
\end{equation}
allowing us to modulate the conditional vector to the noise vector as $\vv_{fused} = \vv_t + \boldsymbol{\gamma} \odot( \mU \vc) + \boldsymbol{\beta} $. Then the fused vector together with  the sinusoidal time encoding, as inputs to an MLP for denoising, as shown in Fig. (\ref{deoiser}).

\vspace{+2mm}
\noindent \textbf{Training.} Following the optimization of the objective function, we focus on training the parameters $\theta$ within the parameterized denoiser $\vphi_{\theta}(\cdot)$, adhering to the DDPM paradigm in our training and parameterization process. The detailed procedure is outlined as follows:

We initially select user interaction vectors $\vx_0$ from the training dataset. Subsequently, a fraction of the interaction history is randomly masked with a 50\% probability, serving as the condition $\vc$ as reported in \cite{diffrec}. The vectors are then transformed into the spectral domain, where noise is added to the initial vector at time step $t$. The resulting noisy vector, in conjunction with the randomly masked condition $\vc$, is fed into the denoising network for denoising. The denoising network $\vphi_{\theta}(\cdot)$ is trained by minimizing the mean squared error (MSE) loss between the generated vector and the clean target vector.

The key idea of this method is described in Algorithm ~\ref{alg:training} and the output parameters of the denoising network.


\vspace{+2mm}
\noindent \textbf{Sampling.} Once the model is trained, we need a method to generate samples for inference. The simplest idea is to utilize our trained model, $\vphi_{\theta}(\vv_t,\vc,t)$, to estimate the noise present in the vector that represents the true interactions. Specifically, whenever we want to move from noise level $t$ to noise level $t - 1$, we can input the noise vector $x_t$ into the model to obtain an estimate of the noise-free interaction vector, $\hat x_0$, and then re-add noise to reach level $t - 1$. This idea is summarized in Algorithms ~\ref{alg:naive_sampler}.  It is worth noting that inspired by \cite{graph-diff}, we use all the historical interactions of users in the test set as the conditional vector $\vc$, without masking. We also add noise to this vector to obtain the noisy vector from which we sample to recover.

To collaborate with the Classifier-free training paradigm, we estimate two clear signals, one from a conditional model and the other from an unconditional model, and perform a weighted average of the two signal estimates. The guidance scale (s) is used to regulate the influence of the conditional signal, where a larger scale produces results that are more consistent with the condition, while a smaller scale produces results with less association.  We have used an unconditional estimate of $s=0.02$ for guidance.



\begin{table*}[!ht]
\scriptsize
\setlength{\abovecaptionskip}{0.05cm} 
\setlength{\belowcaptionskip}{0cm} 
\caption{Overall performance comparison: We highlight \firstone{best} and \secondone{second-best} values for each metric.}
\label{tab:diffrec_clean}
\setlength{\tabcolsep}{2.0mm} 

\resizebox{\textwidth}{!}{%
\begin{tabular}{lcccc|cccc|cccc}
\toprule
& \multicolumn{4}{c|}{\textbf{Amazon-book}} & \multicolumn{4}{c|}{\textbf{Yelp}} & \multicolumn{4}{c}{\textbf{ML-1M}} \\
\textbf{Methods} & \textbf{R@10} & \textbf{R@20}  & \textbf{N@10}  & \textbf{N@20}  & \textbf{R@10}  & \textbf{R@20}  & \textbf{N@10}  & \textbf{N@20}  & \textbf{R@10}  & \textbf{R@20}  & \textbf{N@10}  & \textbf{N@20}  \\ 
\midrule
MF & 0.0437 & 0.0689 & 0.0264 & 0.0339 & 0.0341 & 0.0560 & 0.0210 & 0.0276 & 0.0876 & 0.1503 & 0.0749 & 0.0966 \\
LightGCN & 0.0534 & 0.0822 & 0.0325 & 0.0411 & 0.0540 & 0.0904 & 0.0325 & 0.0436 & 0.0987 & 0.1707 & 0.0833 & 0.1083 \\
CDAE & 0.0538 & 0.0737 & 0.0361 & 0.0422 & 0.0444 & 0.0703 & 0.0280 & 0.0360 & 0.0991 & 0.1705 & 0.0829 & 0.1078 \\
MultiDAE & 0.0571 & 0.0855 & 0.0357 & 0.0442 & 0.0522 & 0.0864 & 0.0316 & 0.0419 & 0.0995 & 0.1753 & 0.0803 & 0.1067 \\
MultiDAE++ & 0.0580 & 0.0864 & 0.0363 & 0.0448 & 0.0544 & 0.0909 & 0.0328 & 0.0438 & 0.1009 & 0.1771 & 0.0815 & 0.1079 \\
MultiVAE & 0.0628 & 0.0935 & 0.0393 & 0.0485 & 0.0567 & 0.0945 & 0.0344 & 0.0458 & 0.1007 & 0.1726 & 0.0825 & 0.1076 \\
CODIGEM & 0.0300 & 0.0478 & 0.0192 & 0.0245 & 0.0470 & 0.0775 & 0.0292 & 0.0385 & 0.0972 & 0.1699 & 0.0837 & 0.1087 \\
DiffRec & \text{0.0895} & \text{0.1010} & \text{0.0451} & \text{0.0547} & \text{0.0581} & \text{0.0960} & \text{0.0363} & \text{0.0478} & \text{0.1058} & \text{0.1787} & \text{0.0901} & \text{0.1148} \\

LinkProp & \text{0.1087} & \text{0.1488} & \text{0.0709} & \text{0.0832} & \text{0.0604} & \text{0.0980} & \text{0.0370} & \text{0.0485} & \text{0.1039} & \text{0.1509} & \text{0.0852} & \text{0.1031} \\
BSPM & \text{0.1055} & \text{0.1435} & \text{0.0696} & \text{0.0814} & \text{0.0630} & \secondone{0.1033} & \text{0.0382} & \text{0.0505} & \text{0.1107} & \text{0.1740} & \text{0.0838} & \text{0.1079} \\
 \text{Giff} &\secondone{ 0.1109} & \secondone{ 0.1521} & \secondone{ 0.0733} & \secondone{ 0.0865} & \firstone{ 0.0639}& { 0.0992}& \firstone{ 0.0397}& \secondone{ 0.0520}  & \secondone{ 0.1108}& \secondone{ 0.1977}&\secondone{ 0.0952}& \secondone{ 0.1176}\\
\midrule

    \text{S-Diff}  &\firstone{ 0.1155} & \firstone{ 0.1604} & \firstone{ 0.0746} & \firstone{ 0.0876} & \secondone{ 0.0635}& \firstone{ 0.1075}& \secondone{ 0.0392}& \firstone{ 0.0561}  & \firstone{ 0.1277}& \firstone{ 0.2018}& \firstone{ 0.0970}& \firstone{ 0.1225}\\
\bottomrule
\end{tabular}
}
\vspace{-4mm}
\end{table*}

\vspace{-4mm}
\section{Experiment} 
\label{Sec:Experiment}
\subsection{Settings}
We conducted experiments on three publicly available datasets and used A800s and Tensorflow for training and inference.  In this work, we set the batch size to 100, and the learning rate to 1e-4, and do not enable weight decay. We train the model for a maximum of 1,000 epochs. In the temporal dimension, inspired by the successful application of diffusion models in collaborative filtering \cite{diffrec}, we set $T=5$, $\tau=1$, and $t_k$ is configured as a linearly spaced sequence between $[0,5]$. Consequently, $t = \tau t_k/T \in [0, 1]$.

\vspace{+2mm}
\noindent  \textbf{Baselines.} We compare our model  with other widely used collaborative filtering methods.

\noindent \ding{182} \textbf{MF} \cite{rendle2009bpr} is one of the most famous collaborative filtering methods, which optimizes the BPR loss using matrix factorization.

\noindent \ding{183}\textbf{LightGCN} \cite{he2020lightgcn} introduces graph convolutional networks (GCNs) without nonlinear functions into collaborative filtering, iteratively aggregating neighborhood information to learn user and item representations.

 \noindent \ding{184} \textbf{CDAE} \cite{wu2016collaborative} trains an autoencoder (AE) to recover user possible true preferences from degraded vectors.

\noindent \ding{185} \textbf{MultiDAE and MultiDAE++} \cite{liang2018variational} utilizes random masking to break interaction signals and trains the AE to recover this signal.

\noindent \ding{186} \textbf{MultiVAE} \cite{liang2018variational} uses variational autoencoders (VAEs) to model the generation process of interaction signals in collaborative filtering and approximates the posterior distribution with an encoder.

\noindent \ding{187} \textbf{CODIGEM} \cite{walker2022recommendation} is the first generative model utilizing diffusion processes, which uses multiple autoencoders to simulate the reverse generation process and uses the predicted vectors for Top-k recommendations.

\noindent \ding{188} \textbf {DiffRec} \cite{diffrec} refines interaction vectors by mapping them to latent spaces further, and it also proposes weighting the interaction sequences based on timestamps.

\noindent \ding{189} \textbf{LinkProp\cite{linkprop} and BSPM\cite{BSPM}} are examples of graph signal processing techniques applied to collaborative filtering that perform better than embedding-based models on large sparse datasets. BSPM also simulates the sharp process's thermal equation, providing a great inspiration for this work.

\noindent \ding{190} \textbf{Giff} \cite{graph-diff} is a specific case of spatial diffusion, smoothing user signals with graph filters and restoring collaborative signals with corresponding graph denoising operators, achieving outstanding results.

\begin{table}[t]
\footnotesize
  \captionsetup{skip=1pt}
  \caption{The statistics of three datasets.}
  \begin{tabular}{cccccl}
    \toprule
    Dataset & \# of users & \# of items & \# of interactions & \# sparsity \\
    \midrule
    MovieLens-1M & 5,949 & 2,810 & 571,531 & 96.6\% \\
    Yelp & 54,574 & 34,395 & 1,402,736 & 99.93\% \\
    Amazon-Book & 108,822 & 94,949 & 3,146,256 & 99.97\% \\
    \bottomrule
  \end{tabular}
  \label{table:datasets}
  \vspace{-5mm}
\end{table}

\noindent \textbf{Datasets.}
To ensure a fair comparison, as shown in Tab. \ref{table:datasets}, we utilize identical preprocessed and partitioned versions of three public datasets: MovieLens-1M, Yelp, and Amazon-Book, with their respective statistics detailed in Table \ref{table:datasets}. Each dataset is split into training, validation, and testing subsets in a 7:1:2 ratio. The validation set determines the optimal epoch for each training method, while the testing set is employed for hyperparameter tuning and final result extraction.

In assessing the top-K recommendation efficacy, we present the average Recall@K and NDCG@K. Recall@K gauges the proportion of relevant items recommended within the top-K list, whereas NDCG@K is a ranking-sensitive metric that assigns higher scores to pertinent items positioned higher in the recommendation list.

\vspace{-2mm}
\subsection{Performance} 
\noindent \textbf{Main Results.}
In Table \ref{tab:diffrec_clean}, we recorded the results for Recall and NDCG metrics. All the results were obtained by averaging over 10 runs. The  results lead us to the following conclusions:

\noindent \ding{172} S-Diff consistently demonstrates significant advantages over all competitors in the recommendation systems, regardless of the dataset or evaluation metrics.

\noindent \ding{173} S-Diff exhibits more stable performance compared to other generative models, particularly in metrics like Recall@20, suggesting that anisotropic diffusion on the spectrum effectively preserves users' global preferences, leading to more stable Top-K recommendations.

\noindent \ding{174} The observed improvement across various metrics over previous graph filters indicates that S-Diff's enhanced performance is attributed to the sophisticated multi-step denoising process, which facilitates the optimization of the variational lower bound's robustness, thereby conferring it with robust spectral expressiveness.
\begin{figure}[!tb]
\centering
\caption{The model's inference time cost for different orders of magnitude of users and items.}
\includegraphics[width=0.45\textwidth]{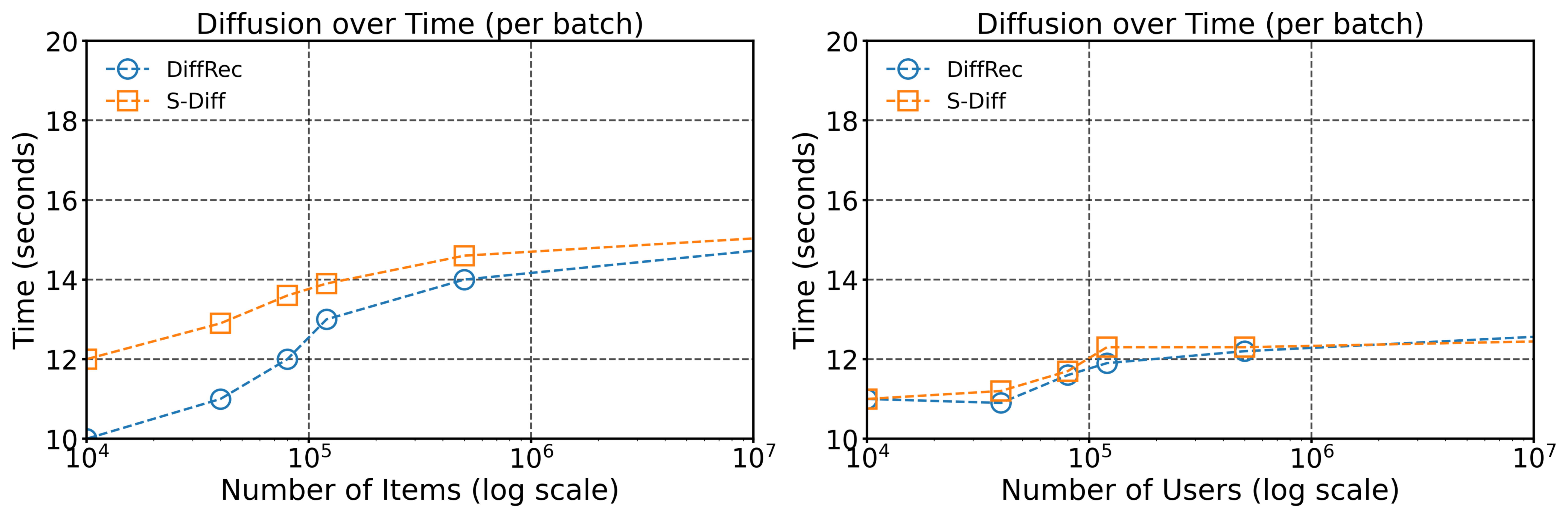}
\label{time-cost}
\vspace{-3mm}
\end{figure}
\noindent \textbf{Time Cost.}
Comparing DiffRec \cite{diffrec}, the well-known spatial diffusion paradigm, with the addition of Gaussian noise in the spatial domain and subsequent reverse restoration. In Figure \ref{time-cost}, we present the time cost of two diffusion models for inferring a single interaction vector. Following the methodology in \cite{cf-diff}, we gradually increase the dimension of the collaborative training interaction matrix and record the time taken for one batch of training inference (batch size = 100). Since the diffusion model needs to generate the click probabilities for all items during the decoding process, it is highly sensitive to the dimensionality of items. As the item dimension increases, the time cost rises significantly. At the same time, the diffusion model is relatively insensitive to the dimensionality of users, even though user nodes are used as stepping stones in the calculation of item-item adjacency matrices. Additionally, S-Diff may initially incur greater costs as it requires additional mapping and remapping, but as the dimensions increase, the time costs converge. Optimizing time costs is also part of future work.

\begin{table}[t]
\scriptsize
\setlength{\tabcolsep}{1pt}
\centering
\caption{Frequency-dependent Noise Parameters 
}
\vspace{-2mm}
\label{table:average_results}
\begin{tabular}{lcccccc}
\toprule

& \multicolumn{3}{c}{\textbf{ML-1M}}
& \multicolumn{3}{c}{\textbf{Yelp}} \\

\cmidrule(lr){2-4} \cmidrule(lr){5-7}
& R@10  & N@10  & log(SNR)  &R@10 & N@10  & log(SNR)   \\
\midrule
 DDPM in Spectral~
& 0.0998   & 0.901      & 5.25   &0.0595  &0.0374     &7.74  \\

S-Diff-VE
& \underline{0.1207}   & \underline{0.0942}  & \underline{8.90}    & \underline{0.0610} & \underline{0.0384} &  \underline{15.89} \\
S-Diff-VP
& \textbf{0.1255}   & \textbf{0.0970}  & \textbf{12.73}    & \textbf{0.0635} & \textbf{0.0392} &  \textbf{21.55} \\
\bottomrule
\vspace{-7mm}
\end{tabular}
\vspace{-3mm}
\end{table}
\begin{figure}[!tb]
\centering
\caption{We removed the FiLM layer successively, replaced it with concat, and obtained w.o. FiLM, and then used conditional-guidance training to conduct ablation experiments on the denoiser network.}
\includegraphics[width=0.45\textwidth]{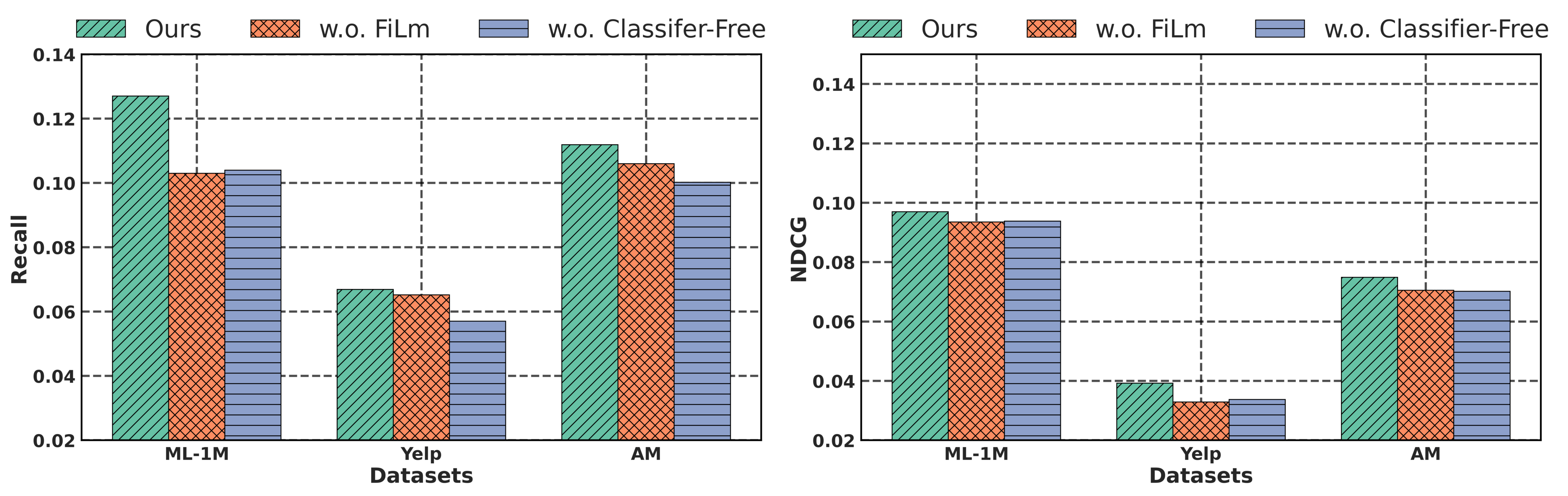}
\label{abiation}
\vspace{-6mm}
\end{figure}

\vspace{-2mm}
\subsection{Ablation}
\noindent \textbf{Spectral Parameters.}
In Tab. \ref{table:average_results}, we examine the setup in Equation (\ref{spectralalpha}), with the aim of understanding the impact of frequency-dependent anisotropic noise on model performance. To this end, we employ the classical DDPM \cite{ddpm} noise schedule in the spectral domain, which we call DDPM in Spectral, where we set $\alpha$ and $\sigma$ as constant factors independent of $\lambda$. Given that we configure the square of our diffusion parameters, to sum up to 1 (see Sec. \ref{vpdiff}), adhering to what is known as variance-preserving Diffusion (VP-Diff), we further delve into the effects of the variance-exploding Diffusion (VE-Diff) \cite{song2020denoising} paradigm in our experiments. Specifically, in VE-Diff, the control of noise $\boldsymbol{\sigma}$ is no longer constrained by $\boldsymbol{\lambda}$ and increases uniformly over time steps, as illustrated in the table. The experimental outcomes suggest two key insights:

\noindent \ding{182} The success of the proposed S-Diffusion (S-Diff) paradigm lies in its spectral diffusion approach, which connects with the spectral domain. When employing isotropic diffusion patterns solely in the spectral domain, the performance is akin to that of spatial domain diffusion, highlighting the significance of spectral considerations.

\noindent \ding{183} S-Diff-VP, in comparison to variance exploding Diffusion models, achieves a higher SNR, aligning with our intuition and leading to superior performance. This underscores the importance of a favorable SNR in the recovery of collaborative signals within recommendation tasks, demonstrating that an element-wise controlled noise is crucial for enhancing the precision of collaborative signal restoration.

\vspace{+1mm}
\noindent \textbf{Denoise Network $\vphi_{\theta}(\cdot)$.} In Fig. \ref{abiation}, we conduct ablation experiments on a noise reduction network, successively replacing the FiLM layer with vector concatenation and modifying the training method to classifier guidance to assess the impact of different approaches.

\begin{figure}[!tb]
\centering
\caption{Sensitivity of the truncated dimension $K$ in matrix decomposition: We examined the impact of the dimension of approximate matrix decomposition on performance across different datasets.}
\includegraphics[width=0.47\textwidth]{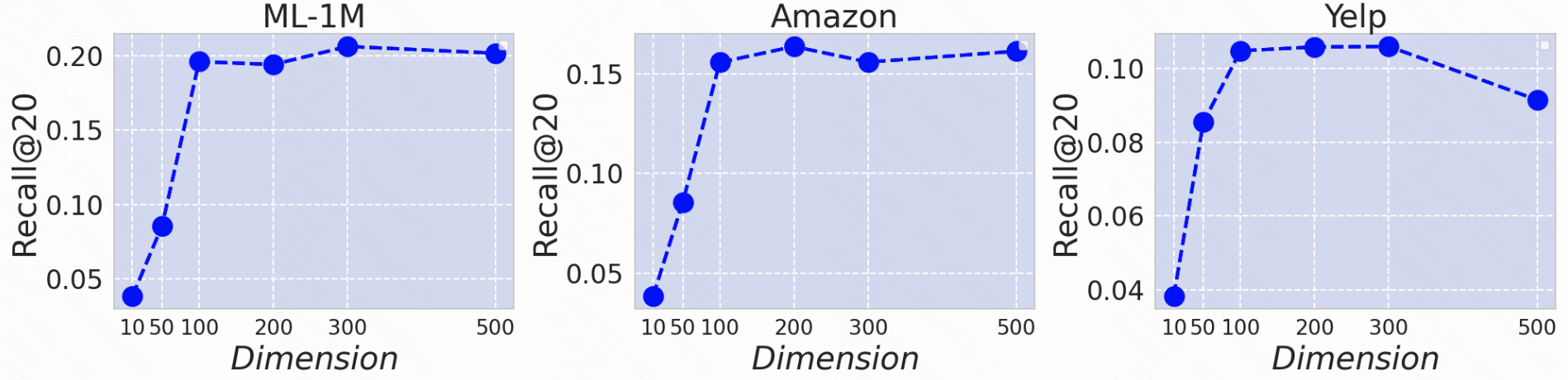}
\label{dim}
\vspace{-5mm}
\end{figure}
\begin{figure}[!tb]
\centering
\caption{Parameter sensitivity of the bounded noise schedule: We jointly evaluated the impact of the parameters $\alpha_{\min}$ and $\sigma_{\max}$ on the recall rate.}
\includegraphics[width=0.47\textwidth]{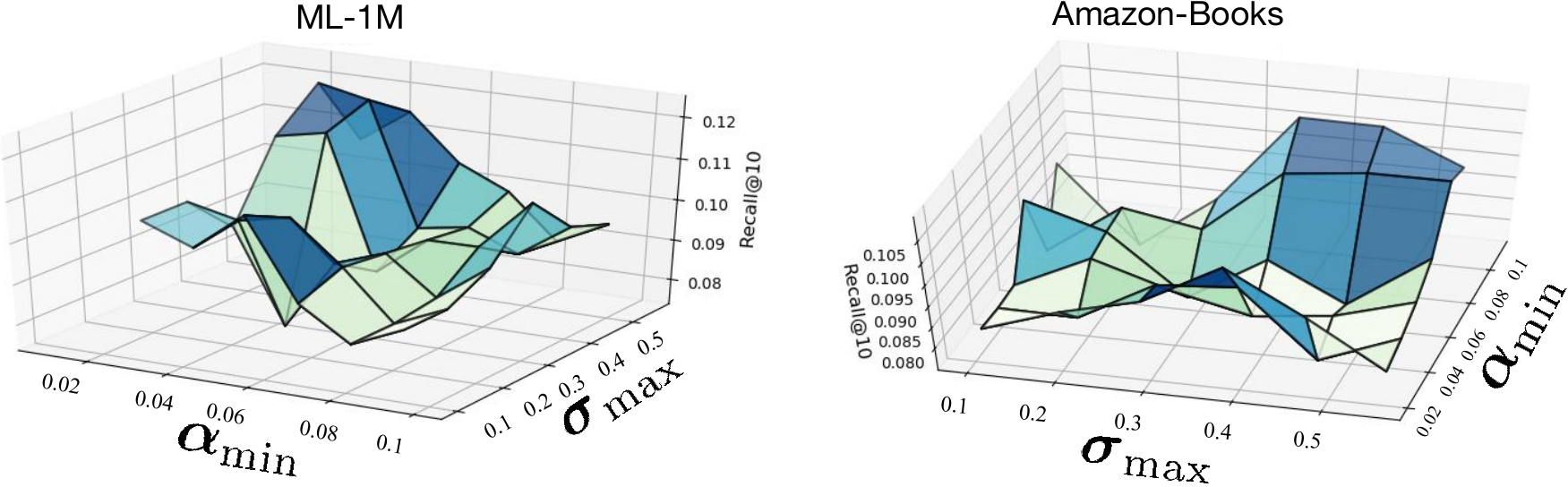}
\label{para}
\vspace{-5mm}
\end{figure}

\vspace{-2mm}
\subsection{Sensitivity}
\noindent \textbf{Dimension of spectral decomposition.}
In Fig. \ref{dim}, we show the influence of spectral decomposition dimension on the performance of the spectral domain diffusion model when it is used for recall. It is worth noting that on the three data sets, we obtain good benefits when we take 200-dimensional spectral decomposition, which is in line with the practice of past researchers.

\noindent \textbf{Boundedness of the diffusion noise schedule.} In Fig.\ref{para}, we discuss the impact of threshold  parameters on the spectral diffusion model. When we adopt a fixed time step, we impose constraints on the boundedness of $\alpha_t$ and $\sigma_t$: the  preservation term $\alpha_{min}$ and the noise upper bound $\sigma_{max}$. We conduct parameter sensitivity experiments on the MovieLens and Amazon-Book datasets, indicating that a smaller lower bound for feature preservation $\alpha_{min} \in [0,0.1] $ and a larger upper bound for variance $\sigma_{max} \in [0.4,0.5]$ contribute to improved performance. Interestingly, on the Amazon-Book dataset, it is necessary to retain smaller eigenvectors for diffusion to achieve the best results.

\vspace{-3mm}
\section{Related Works}
\label{Sec:Related}
In the introduction, we briefly discussed the related work. Here, we delve deeper into the relevant methods.

On one hand, diffusion models have been employed in recommendation systems \cite{diffrec, dreamrec, walker2022recommendation, difaug}. In \cite{walker2022recommendation}, the DDPMs paradigm was first used in collaborative filtering to recover large-scale interaction matrices. Diffrec \cite{diffrec} improved the accuracy of generative collaborative filtering by using a diffusion model with time-step-guided diffusion and encoding in a latent space. Similarly, methods like DreamRec embed interaction vectors into low-dimensional spaces. It's worth noting that spectral diffusion models can be seen as a particular case of diffusion in the latent space, reducing the computational complexity of high-dimensional diffusion. 
The ability of diffusion models for recommendation to capture common user preferences is a question worth considering, as it forms the basis of collaborative filtering's enduring success \cite{He2017Neural}. CF-Diff \cite{cf-diff,cfdiff2} addresses this by pre-computing user's multi-hop neighborhood information and encoding this information into the conditional of a conditional diffusion model, which may have suboptimal scalability. Giff \cite{graph-diff}, on the other hand, makes encouraging progress by first breaking down user interaction vectors and then recovering them through graph propagation. Building upon these foundations, we extend the definition of recommendation diffusion models defined on graph spectra, allowing for more flexible parameterization.

Naturally, this anisotropic parameterization process reminds us of the achievements graph filters have made in the past. Among the most notable practices, LightGCN \cite{he2020lightgcn} aggregates neighborhood information using multi-layer linear encoding, considered an example of parameter-free graph diffusion. From a spectral theory perspective, it's often viewed as a low-pass filter. Poly-CF \cite{qin2024polycf} enhances the impact of different frequencies on collaborative filtering by employing adaptive spectral graph filtering. SGFCF \cite{howpowerful} additionally demonstrates that collaborative signals at different frequencies contain varying levels of noise, transforming collaborative filtering into a parameterization problem—how to design an adaptive filter to recover the true signal. These works provide crucial theoretical insights, prompting us to experiment by introducing noise at different scales on frequencies corresponding to different eigenvalues and recovering from it.

\vspace{-2mm}
\section{Conclusion}
\label{Sec:Conclusion}
In this paper, we introduce a spectral domain diffusion paradigm for collaborative filtering in recommendation systems, which implicitly captures users' shared preferences. We investigate the theoretical and empirical capabilities of this paradigm, demonstrating its advantages across diverse datasets, and enriching the practice of diffusion models in the recommendation community.

\bibliographystyle{ACM-Reference-Format}
\bibliography{wsdm24/wsdm}

\clearpage 
\appendix
\section{Appendix}
\begin{table}[ht]
    \centering
    \caption{Symbol table.}
    \small
    \begin{tabular}{l|p{6cm}} 
        \toprule
        \textbf{Symbol} & \textbf{Description} \\
        \midrule
        $\vx_0$ & Initial interaction vector drawn from  user data distribution$q(\vx_0)$. \\
        \midrule
        $\vx_t$ & Noisy interaction vector at time step $t$. \\
        \midrule
        $\boldsymbol{\alpha_t}$ & Time-dependent scaling factor. \\
        \midrule
        $\boldsymbol{\sigma_t}$ & Time-dependent noise scale. \\
        \midrule
        $\boldsymbol{L}$ & Graph Laplacian matrix. \\
        
        \midrule
        $\boldsymbol{A}$ & Adjacency matrix of the graph. \\
        \midrule
        $\mU$ & Orthonormal basis of eigenvectors from spectral decomposition of  $\boldsymbol{L}$. \\
        \midrule
        $\mD_t$ & The diagonal matrix with eigenvalues, $\mD_t = \operatorname{diag}\{-t d_1, \ldots, -t d_{|\mathcal{I}|}\}$, satisfies $(-\boldsymbol{L} t) = \mU \mD_t \mU^\mathrm{T}$.\\
        \midrule
        $\mC_t$ & The forward graph heat diffusion operator, $\mC_t$, satisfies $\mC_t = e^{ -\boldsymbol{L} t} = \mU \boldsymbol{\Lambda}_t\mU^\mathrm{T}$. \\
        \midrule
        $\boldsymbol{\Lambda}_t$ & The diagonal matrix obtained from the spectral decomposition of $\mC_t$.\\
        \midrule
        $\boldsymbol{\lambda}_t$ &The eigenvectors formed by different dimensions of  $\boldsymbol{\Lambda}_t$, also implying the high-frequency and low-frequency information of the graph.\\
        \midrule
        $\bm{\eps}_t$ & Gaussian noise, $\bm{\eps}_t \sim \mathcal{N}(\bm{0}, \bm{I})$. \\
        \midrule
        $\vc$ & User's historical interaction data. \\
        \midrule
        $\vphi_\theta$ & Neural denoiser parameterized by $\theta$. \\
        \midrule
        $\hat{\vx}_0$ & Final denoised preference vector. \\
        \midrule
        $\vv_0$ & Latent spectral vector, $\vv_0 = \mU^{T} \vx_0$. \\
        \midrule
        $\vv_t$ & Noisy latent spectral vector at time step $t$. \\
        \midrule
        $s$ & Strength of unconditional guidance. \\
        \midrule
        $p_{\text{uncond}}$ & Probability of unconditional sampling. \\
        \bottomrule
    \end{tabular}
    \label{tab:symbols}
\end{table}

\end{document}